# *Stegosaurus* chirality

R. P. Cameron[1,*], J. A. Cameron[1], and S. M. Barnett[1]
[1]School of Physics and Astronomy, University of Glasgow, Glasgow G12 8QQ, United Kingdom.
*robert.cameron@glasgow.ac.uk

**Introduction**
Most living things appear rather symmetrical: the external human form, for example, has a plane of mirror symmetry to good approximation. Snails, flounders, narwhals, crossbills, fiddler crabs and twining vines are members of the short but fascinating list of living things known instead to defy mirror symmetry by exhibiting exterior chirality[1]. The study of this symmetry breaking lies at the cutting edge of developmental and evolutionary biology[2,3]. We have recently extended the aforementioned list[4] by adding one of the most recognisable genera of dinosaurs: *Stegosaurus*[5]. Here we summarise our research to date into *Stegosaurus* chirality.

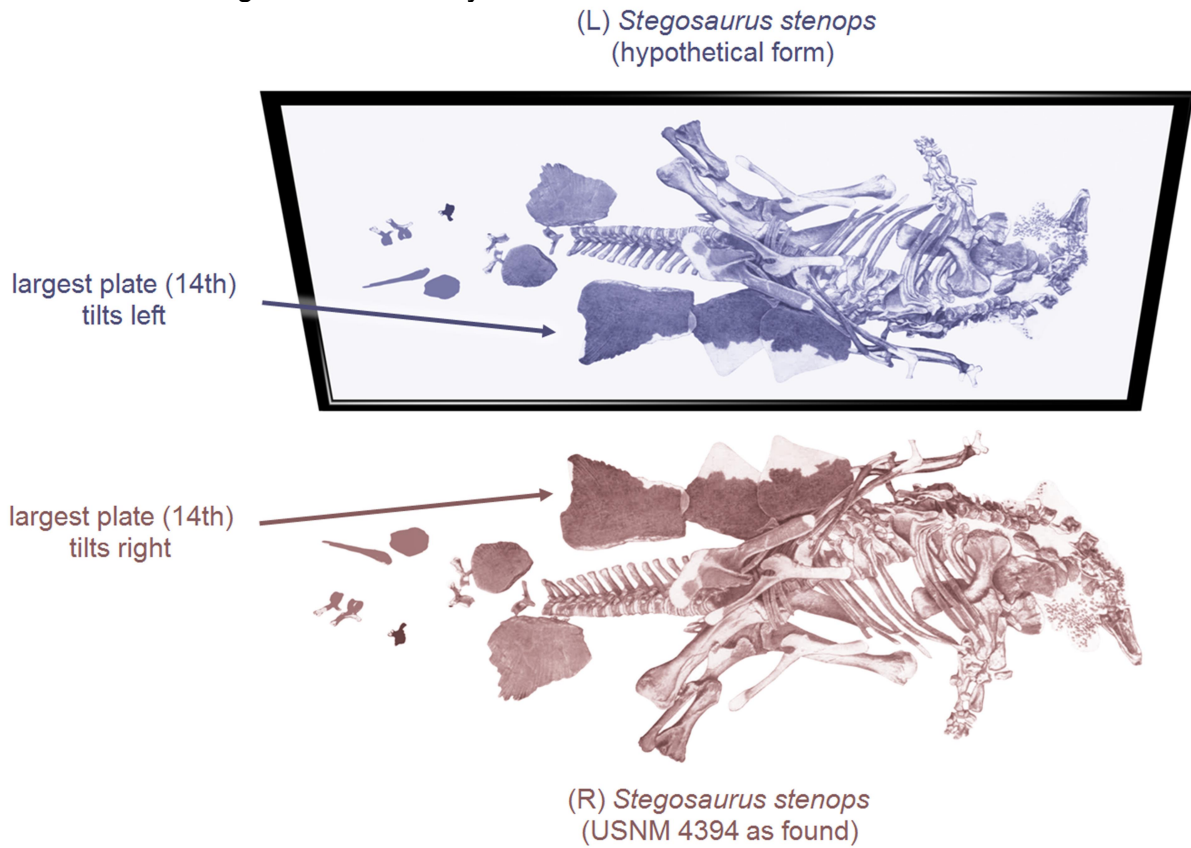

**Figure 1.** The arrangement of *Stegosaurus*'s plates differs from its mirror image and is therefore chiral[1], as highlighted here for the largest plate in particular of the *Stegosaurus stenops* holotype USNM 4939: this specimen was of the (R) rather than the (L) form. Adapted from [8].

The currently favoured arrangement of *Stegosaurus*'s plates was put forward by Lucas[6-8] and sees them mounted in two staggered rows along the animal's back. There are, in fact, *two* conceivable forms consistent with this basic description: if the largest plate in particular tilts to the right we have an (R) *Stegosaurus*, if it tilts to the left we have an (L) *Stegosaurus* instead[4]. That *Stegosaurus* exhibited exterior chirality is beyond reasonable doubt: many of the plates are manifestly chiral by themselves and no two plates of the same size and shape have been found for an individual[8-10]. Plates have been correlated between individuals, however[8]. The exterior chirality of *Stegosaurus* may be of particular interest to evolutionary biologists, as exterior chirality is especially rare amongst modern-day reptiles and also birds[2,3].

**Results**

A survey based upon the well-preserved remains of four *Stegosaurus* specimens of the same species (*stenops*), comparable ontogeny (subadult to old adult), same epoch (Upper Jurassic) and similar location (North America) is consistent with our hypothesis that the (R) form was genetically favoured: the largest plate of USNM 4394 (Garden Park, Colorado[8,11]; see Figure 1), USNM 4714 (Como, Wyoming[8]), DMNS 2818 (Garden Park, Colorado[12]) and NHMUK R36730 (Red Canyon Ranch, Wyoming[13]) appears to have tilted to the right in each case. Perhaps the (L) form existed less frequently, due to mutation: sinistral individuals are sometimes born within predominantly dextral populations of snail[2-4], for example.

The exterior chirality of *Stegosaurus* has not been described explicitly before by others. More importantly, the need to distinguish a specimen from its distinct, hypothetical mirror-image form does not appear to have been recognised: Felch used the words "*left*" and "*right*" interchangeably in his correspondence with Marsh[8], thus giving a chirally ambiguous description of USNM 4394; Gilmore exhibited USNM 4394 together with "*a large mirror*", thus displaying an individual that never existed[8,14]; Carpenter described DMNS 2818 together with a drawing of an (L) *Stegosaurus stenops*[12], which is at odds with the (R) assignment described above. The world's most complete *Stegosaurus* specimen NHMUK R36730 appears to be on display at present with its largest plate orientated incorrectly such that the plate's right surface is facing left and its left surface is facing right[13]…

**Discussion**

The most popular hypothesis for the function of the plates is that they acted primarily as display structures, perhaps to ward off predators, to aid in identification or as a means of attracting mates[8,12,15]. The possibility that males and females had different plate morphologies seems to tie in particularly with well with the latter ideas[16]. The high degree of vascularisation evident in the plates[17] has led in particular to claims that they could "*blush*" so as to embellish their appearance[12,15]. The chirality of the plates makes the idea that they served as display structures all the more plausible, for it is integral to their appearance. We note in particular that two staggered rows of plates might give a more substantial lateral profile than two parallel rows of plates, for example, as the latter might yield visible gaps where the former has none. Chirality might have been nature's way, therefore, of granting *Stegosaurus* a body-length sail of maximised apparent area, without overly restricting the animal's movement[6,12]. In this role there is no obvious reason to prefer the (R) or the (L) form. The dominance of the (R) form suggested above for *Stegosaurus stenops* then seems natural: with no obvious need for variation between forms, consistency might prevail, as is the case for many modern-day living things exhibiting exterior chirality[18]. We note with interest that the first description in print of the currently favoured arrangement of *Stegosaurus*'s plates is rather close in spirit to our observations: Lucas noted that the plates "*were placed far enough apart to permit freedom of motion, and appear to have been arranged alternately and not in pairs*"[19].

Whether an (L) *Stegosaurus* ever existed remains to be seen. Lucas once described *Stegosaurus* as being "*among the most singular of all known animals, singular even for Dinosaurs*"[19]. Perhaps he was only half correct.

**Materials and method**

For USNM 4934 and USNM 4714 we based our investigation primarily upon an original copy of Gilmore's seminal monograph[8]. For DMNS 2818 we based our investigation upon Carpenter's paper[12] and, in particular, communications with Evan Saitta, who has observed the specimen first-hand[20]. For NHMUK R36730 we based our investigation upon documentation by Siber and his team[13] as well as first-hand observations by R. P. C. and J. A. C. of the specimen in the Natural History Museum, London in 2015. Our method was simply to assign an (R) or an (L) chirality to each of these specimens by careful inspection of the aforementioned evidence.

Well-preserved *Stegosaurus* remains are exceedingly rare and our modest survey has already exhausted the supply of articulated remains available to us at present for *Stegosaurus stenops*. In principle it should be possible to assign an (R) or an (L) chirality to a specimen by extrapolating from the chirality of but one of the specimen's plates (see Figure 2). We attempted this approach but found it to be unsatisfactory in practice as it requires that the precise location of the plate in the series be known and little is certain in this regard. Let us be clear, however, that the position and character of the *largest* plate and thus our definition of (R) and (L) forms is well-established for *Stegosaurus stenops* at least. We were unable to assign chiralities to specimens of any other species with confidence, again owing to the lack of sufficiently well-preserved remains. *Stegosaurus* chirality should be borne in mind by those lucky enough to make new finds.

*Stegosaurus stenops* cervical plate

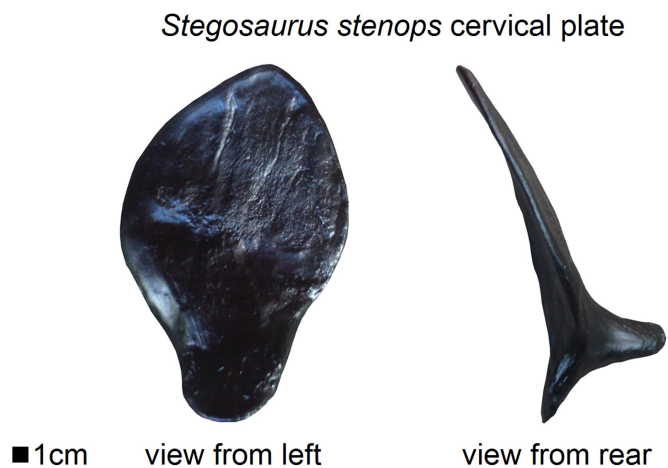

■1cm     view from left     view from rear

**Figure 2.** A cast of a *Stegosaurus stenops* cervical plate, which is manifestly chiral by itself and appears to have tilted to the left rather than to the right. The chirality seen here is not likely to be an artifact of postmortem distortion, for example: the "*asymmetrical*" bases of cervical plates in particular clearly and consistently match the contours of nearby vertebra[8,9].

**Notes of potential interest**
Lull, who advocated an (incorrect) symmetrical arrangement of two parallel rows of plates[7,9,10,21], appeared to do so primarily because, in his view, the "*fact that in no other reptile the dermal elements alternate seems too weighty an argument to be lightly dismissed*"[9]. This might be interpreted as a failure to appreciate the possibility of exterior chirality: it is true indeed that exterior chirality is rare in living things, particularly amongst modern-day reptiles and indeed birds[3], but it is certainly not unprecedented.

It is not yet known how to assign genders to *Stegosaurus* remains and we have made no attempt to do so, although the sexual dimorphism identified recently for *Hesperosaurus mjosi*[16] suggests, perhaps, that different 'species' of *Stegosaurus* may in fact have been different genders of the same species.

It has been suggested that the 'asymmetry' of the plates' arrangement may have been absent from juveniles, the dermal spikes of the rhinoceros iguana *Cyclura cornuta* having been cited by way of example[10]. There is some evidence to suggest that juveniles did not have had plates at all[22].

There does not appear to be a collective noun particular to *Stegosaurus*, although the word 'herd' is sometimes employed. We suggest that a collection of more than one *Stegosaurus* might be referred to henceforth as a 'handful' of *Stegosaurus*, to reflect the exterior chirality and apparent rarity of the animals.

**Acknowledgments**
This work was supported by the Engineering and Physical Sciences Research Council (EP/M004694/1 and EP/I012451/1) and the Royal Society.

We thank Revinder Chahal for her swift and helpful correspondences; Roslyn Taplin for her invaluable advice with regards to the drafting of this manuscript and Evan Saitta for lending us his keen insights, in particular with regards to DMNH 2818.